\documentclass[copyright,creativecommons]{eptcs}
\providecommand{\event}{FMAS 2026} 

\usepackage{iftex}

\ifpdf
  \usepackage{underscore}         
  \usepackage[T1]{fontenc}        
\else
  \usepackage{breakurl}           
\fi

\usepackage{amssymb}
\usepackage{amsmath}
\usepackage{amsthm}
\usepackage{graphicx}
\usepackage{framed}

\title{Let's Talk This Out: \\Towards Explanation-based Distributed Goal Satisfaction}
\author{Anton Reinecke \qquad\qquad Sibylle Schupp
\institute{Hamburg University of Technology\\ Hamburg, Germany}
\email{anton.reinecke@tuhh.de \qquad\qquad schupp@tuhh.de}
}
\def\titlerunning{Let's Talk This Out}
\def\authorrunning{A. Reinecke \& S. Schupp}
\begin{document}
	\maketitle

	\begin{abstract}
	Digital systems are embedded in a modular environment of different components with individual goals and settings, which are often installed without central oversight. In this research preview paper we present a framework for communication between components to facilitate compositional problem solving based on explanation patterns and A/G-contracts, without the need for an oracle. Our approach uses partial MaxSMT solving and quantifier elimination to compute counterfactual explanation messages. We illustrate the feasibility of the approach with a proof-of-concept chain topology example, and discuss future challenges when expanding the protocol to more complex topologies.
	\end{abstract}

	\section{Introduction}
		\label{sec:introduction}
		 Cyber-physical systems (CPS) exist in, and interact with, their environment and other CPSs in it. Because of these interactions, software verification methods are incomplete that only take a single component of the greater system into account, without considering its effects on and influences from the environment. The state of the environment and its possible natural evolution, as well as the actions and reactions of other components in it, lead to a state space explosion that makes it infeasible to derive the correct set of actions during development. At the same time, the precise composition of the system and all its components is generally unknown beforehand if we consider modular systems where a user may arrange and replace components dynamically.

		 Take, for example, a modern energy infrastructure utilizing a broad range of decentralized renewable energy sources. The individual generators may come from different producers and editions, but still be connected to the same electricity grid, and supply machines which are often even more diverse in their origins and function. Even if one could plan the whole infrastructure at one point, it is bound to change due to replacements or repairs, or even extreme changes in circumstances such as storms, or unusually long stretches of time without direct sunlight. How can the correct cooperation between all of these components then be guaranteed?

		 We propose a five-step workflow for the generation and use of explanations based on assume-guarantee contracts (A/G-contracts), minimal correction sets (MCS), and quantifier elimination (QE): (1)~a component detects a conflict between its environment assumptions and the target it guarantees; (2)~it identifies a responsible MCS candidate among its environment assumptions; (3)~it computes maximally permissive relaxations of these assumptions via QE; (4)~it sends them as a counterfactual explanation to the component controlling the assumption; and (5)~the receiver either returns a new value satisfying the relaxed constraint and its own target, or a rejection.

		 The explanation generation then needs to be embedded in a communication protocol which terminates and either finds a new solution or proves that none exists. We present the generation of maximally permissive counterfactual explanations between two components, and discuss challenges arising in more complex topologies.

		 We impose the following rules on the system and interactions between components: First, we assume that the system was in a working state where every target was satisfied before a disturbance introduced a fault that now needs to be compensated. Second, each component knows its own interface: which shared variables affect it, which neighbour controls each of them, an exact model of how these variables affect its own target, and the current values of all quantities that affect it but that it does not control (e.g., a measured wind speed). It knows neither the internal models of other components nor the global topology. Learning approaches or separate communications could be used to determine these interactions.

		 \subsection{Related Work}
		 	Interactions between components in a system of systems are an active topic of research, which has been growing in recent years. However, most existing research directions treat it as a game-theoretic problem. Of existing work, we found that Chatterjee et al.~\cite{chatterjee2008} promote a similar concept as we do, i.e., the gradual weakening of environment assumptions through minimal edge removal on a game graph, which mirrors our use of MCSs. Recently, Mu et al.~\cite{mu2026} have explored mechanisms to assign responsibility for outcomes to individual actions of components in a system, which is close in concept to our complaint mechanism. While their approach is post-hoc, Parker et al.~\cite{parker2023} focus on anticipatory responsibility attribution. Finally, some works lie at the intersection of game theory and A/G-contracts, for example Cavezza et al.~\cite{cavezza2020}. Anand et al.~\cite{anand2023} elaborate on the requirements for synthesized assumptions, i.e., sufficiency, actionability, and permissiveness.

		 	Distributed constraint satisfaction and optimization (DisCSP/DCOP)~\cite{yokoo1998,fioretto2018} also coordinates agents that each control some variables of a shared constraint problem. In contrast to that setting, our components pursue heterogeneous local targets and communication is only initiated when one fails to reach it. The only global objective is the fulfillment of all local targets.

		 	A/G-contracts for compositional reasoning were introduced by Misra and Chandy~\cite{misrachandy1981} and Abadi and Lamport~\cite{abadi1993}, and actively used for CPS design since Sangiovanni-Vincentelli et al.~\cite{sangiovannivincentelli2012}. Modern meta-theory and different formalisms are presented by Benveniste et al.~\cite{benveniste2018}. Luckcuck et al.~\cite{luckcuck2026} combine first-order A/G-contracts, compositional reasoning, and runtime monitoring for modular robotic systems. Like ours, their setting is modular and contract-based. However, their contracts are defined at design time and monitored at runtime, whereas we use violated assumptions as the starting point for negotiating new ones at runtime. Their work also provides valuable insight into different network topologies and the challenges one needs to overcome when considering them.

		 	A/G-contracts need to specify environment assumptions for which the defined properties are guaranteed to hold. Therefore, the learning of assumptions which are consistent with all components in a given system is a topic of specific interest. Giannakopoulou et al.~\cite{giannakopoulou2002} introduced the framework of assumption learning, which was refined by Cobleigh et al.~\cite{cobleigh2003} to use Angluin's L* algorithm~\cite{angluin1987}. Several further refinements and applications of this approach exist, such as the extension to probabilistic assumptions by Feng et al.~\cite{feng2011}, and the extension to timed automata by Chen et al.~\cite{chen2023}.
		 	In particular, Frenkel et al.~\cite{frenkel2022} introduce a repair step which is invoked when a bad trace is found during verification.
		 	Diverging from the L*-style assumption learning line, Bobaru et al.~\cite{bobaru2008} instead use a CEGAR-style abstraction refinement.
		 	However, what all approaches have in common is that they take place during development time and require an ``omniscient'' oracle which knows the ground truth. To the best of our knowledge, runtime assumption learning in which information can only be exchanged via peer-to-peer messages has not yet been discussed directly.

		 	Since no component has access to another's internal state, necessary information must be communicated through explicit messages, which need to contain more than a binary (or ternary) verdict. Instead, a message should \textit{explain} why a goal cannot be guaranteed, and do it in a way that is sufficient for the receiver to act on.
		 	In the current research environment, ``explanations'' often refer to explainable AI (see surveys such as Arrieta et al.~\cite{arrieta2020} or Longo et al.~\cite{longo2024}), and other human-centric concepts like trust and post-hoc blame attribution (see for example Shymanski et al.~\cite{shymanski2026}, Wachter et al.~\cite{wachter2018} or Zehtabi et al.~\cite{zehtabi2024}). Our work instead considers machine-to-machine explanations for increased cooperation. A particularly useful family of explanations for this purpose are counterfactual explanations~\cite{lewis1973}, i.e., messages of the form ``If ABC was like this, then XYZ could have been achieved.''
		 	One may ask whether such messages are explanations or simply constraint propagation. We use the term in the technical sense of Fey et al.~\cite{fey2022}: a message is derived from an explanation pattern, it is counterfactual (a minimal change to the belief under which the target holds), and it is actionable for the receiver. In contrast to plain constraint propagation, it carries the reason for the failure, i.e., which assumption is responsible and the maximal range in which it would succeed, rather than just a value or a verdict.

		 	A 2025 survey by Straub et al.~\cite{straub2025} reviewed recent papers concerning explanations in self-adaptive systems. Apart from noticing a recent surge of interest in the topic, they identified the lack of \textit{actionable explanations} as a gap in current research, which is where our paper is located. Other algorithms for self-adaptive systems comprise fault detection and repair, but provide no explanations.

		 	Section~\ref{sec:prelim} introduces background, Section~\ref{sec:explanation} presents the generation of a sufficient, permissive, and actionable counterfactual explanation, Section~\ref{sec:demonstration} demonstrates it on a chain, and Section~\ref{sec:protocol} discusses its embedding in a peer-to-peer protocol.

	\section{Preliminaries}
		\label{sec:prelim}

		In our work we use the explanation pattern introduced by Fey et al.~\cite{fey2022}. Let $(\mathcal{M},\mathcal{B},\mathcal{T})$ be the \textit{explanation pattern} owned by every component in a world, where $\mathcal{M}$ is a model of the component and partial model of its surroundings, $\mathcal{B}$ is a set of constraints defining the (belief about the) current state, and $\mathcal{T}$ is an explanation target -- for example the satisfaction of some specifications. The system tries to derive from $\mathcal{M}$ and $\mathcal{B}$ whether $\mathcal{T}$ is justified, unjustified, or inconclusive.

		An \textit{assume-guarantee contract} (A/G-contract) is a tuple $C = (A, G)$ of assertions called the assumption and the guarantee (cf.~\cite{benveniste2018}). The assumption $A$ generally refers to conditions of the environment. An implementation satisfies $C$ if every environment state that satisfies $A$ also satisfies $G$, i.e., $A \Rightarrow G$.

		Let $F=\{c_1,\dots,c_n\}$ be an unsatisfiable set of clauses. Then a \textit{minimal correction subset} (MCS) is a set of clauses $C \subseteq F$ such that $F \setminus C$ is satisfiable and for every proper subset $C' \subsetneq C$, $F \setminus C'$ is not satisfiable (cf.~\cite{reiter1987, liffiton2008}). An MCS can, for example, be obtained as the complement of a \textit{maximum satisfiable subset} returned by a MaxSMT solver; such a minimum correction set is in particular minimal.

		Given the explanation pattern $(\mathcal{M},\mathcal{B},\mathcal{T})$ and the situation that either $\mathcal{M},\mathcal{B} \models \neg \mathcal{T}$ or $\mathcal{M},\mathcal{B} \models ?\mathcal{T}$, a \textit{counterfactual explanation} is a minimal difference between $\mathcal{B}$ and a $\mathcal{B}'$ such that $\mathcal{M},\mathcal{B}' \models \mathcal{T}$.

		Let $\varphi(\bar{y}) \equiv \exists x.\ \psi(x, \bar{y})$ be a logical formula over the reals and booleans, where $x \in \mathbb{R} \cup \mathbb{B}$ and $\bar{y} \in (\mathbb{R}\cup \mathbb{B})^n$.
		\textit{Quantifier elimination} computes a quantifier-free formula $\varphi'(\bar{y})$, such that
		$\varphi(\bar{y}) \leftrightarrow \varphi'(\bar{y})$
		holds for every assignment of $\bar{y}$ over $\mathbb{R}\cup \mathbb{B}$. Universal quantifiers are eliminated analogously via $\forall x.\,\psi \equiv \neg\exists x.\,\neg\psi$.

	\section{Explanation Generation}
		\label{sec:explanation}
		\begin{figure}[t]
			\centering
			\includegraphics[width=.55\linewidth]{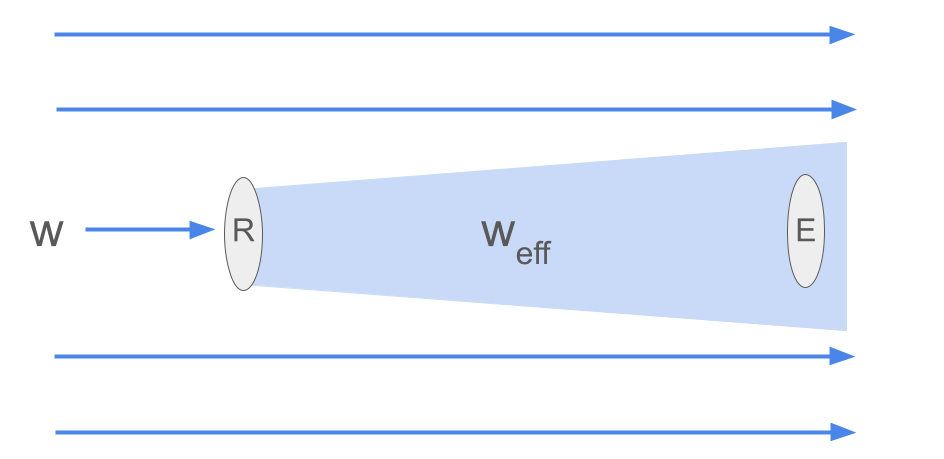}
			\caption{Wind turbines: $E$ is located in the wake of $R$.}
			\label{fig:wake}
		\end{figure}
		In a system, each component holds assumptions about the environment and guarantees which it must uphold. In terms of the previously introduced explanation pattern, the guarantee $G$ corresponds to the explanation target $\mathcal{T}$, while the assumption $A$ is the set of environment states $\mathcal{B}$ for which $\mathcal{M}, \mathcal{B}\models \mathcal{T}$.
		The explanation generation takes place between two individual components, which we will refer to as the explainer ($E$) and the receiver ($R$), whenever $E$ notices that its target cannot be fulfilled under the current believed state of its environment. We consider a setting in which $E$ operates under partial influence of $R$.

		Following the notation of Section~\ref{sec:prelim}, $E$ maintains a model $\mathcal{M}_E$, which is not subject to repair in this work, and a belief $\mathcal{B}_E = \mathcal{B}_{E,H} \cup \mathcal{B}_{E,S}$. $\mathcal{B}_{E,H}$ encodes hard knowledge about $E$'s own state, while $\mathcal{B}_{E,S}$ encodes the soft, shared variables whose values are under $R$'s influence. $\mathcal{B}_{E,S}$ is therefore a subset of $E$'s contract assumptions, specifically the subset of changeable parameters.

		We distinguish the \emph{shared} variables $\bar y$ constrained by $\mathcal{B}_{E,S}$, whose values are controlled by $R$, from all remaining variables $\bar x$. The latter comprise $E$'s own settings, quantities derived via $\mathcal{M}_E$, and uncontrolled quantities whose current value is known from $\mathcal{B}_{E,H}$, e.g., environmental properties.

		\noindent Given an explanation target $\mathcal{T}_E$ such that $\mathcal{M}_E, \mathcal{B}_E \models \neg \mathcal{T}_E$ or $\mathcal{M}_E, \mathcal{B}_E \models ?\mathcal{T}_E$, $E$ computes and communicates a counterfactual explanation as follows.

		\begin{enumerate}
			\item \textbf{Failure detection.} $E$ evaluates if $\mathcal{M}_E, \mathcal{B}_E \models \mathcal{T}_E$. If the target is not justified, explanation generation is initiated.

			\item \textbf{Diagnosis.} $E$ encodes $F = \mathcal{M}_E \wedge \mathcal{B}_{E,H} \wedge \mathcal{T}_E \wedge \mathcal{B}_{E,S}$ as a partial MaxSMT problem, treating clauses from $\mathcal{M}_E$, $\mathcal{B}_{E,H}$, and $\mathcal{T}_E$ as hard, and clauses from $\mathcal{B}_{E,S}$ as soft. Since only $\mathcal{B}_{E,S}$ is eligible for correction, the unsatisfied soft clauses returned by the solver form a \emph{restricted minimal correction subset} $C \subseteq \mathcal{B}_{E,S}$. We write $\bar y_C$ for the shared variables occurring in $C$.

			\item \textbf{Range characterization.} The clauses in $C$ are removed from $F$, leaving $\bar y_C$ unconstrained by beliefs. $E$ then applies quantifier elimination to
			\begin{equation}
				\label{eq:characterization}
				\varphi(\bar y_C) \;\equiv\; \exists \bar x.\, \big(\mathcal{M}_E \wedge \mathcal{B}_{E,H} \wedge \mathcal{T}_E \wedge (\mathcal{B}_{E,S}\setminus C)\big),
			\end{equation}
			yielding a quantifier-free formula over $\bar y_C$ only. Optionally, $\varphi$ is simplified to remove redundant clauses and increase interpretability, e.g., using the Z3 tactic \texttt{propagate-ineqs}.

			\item \textbf{Communication.} The counterfactual explanation $\mathcal{B}'_{E,S} = (\mathcal{B}_{E,S} \setminus C) \cup \{\varphi\}$ is sent to $R$.

			\item \textbf{Response.} $R$ adopts $\varphi$ as a hard constraint in its own belief set $\mathcal{B}_{R,H}$, attempts to satisfy its own target under this new restriction, and returns either a rejection or a new value for $\bar y_C$ satisfying $\varphi$.
		\end{enumerate}

		\noindent Since all uncontrolled quantities have known values, the only freedom left in $\varphi$ lies in $E$'s own settings. A satisfying assignment therefore describes a configuration that $E$ can actually realise, which is the guarantee $E$ needs from $R$. 
		Equation \ref{eq:characterization} encodes this relationship for the explanation formula $\varphi(\bar y_C)$ which must be equivalent to the existence of an assignment of $\bar x$ satisfying the conjunction of model, hard belief, target, and the soft belief without $C$. Quantifier elimination preserves this relationship.
		If uncontrolled quantities were only known up to bounds, the target would have to hold for all admissible values, requiring a universal quantifier in $\varphi$ and a diagnosis step consistent with it. We leave this extension to future work.	
		We chose to use SMT for its expressiveness and mature tool support. Other formalisms such as game graphs (cf.~\cite{chatterjee2008}) have concepts similar to MCSs and should allow an adaptation.

		\paragraph{Example}
		Consider the example shown in Figure~\ref{fig:wake}. Wind turbine $E$ is located downstream of wind turbine $R$, which causes it to miss its energy production target $\mathcal{T}_E$ because of $R$'s wake effect.

		$E$'s model $\mathcal{M}_E$ consists of the physical equations
		\[
		w_{\mathit{eff}} = w \cdot (1 - s), \qquad \mathit{out} = w_{\mathit{eff}} \cdot \eta,
		\]
		where $w = 8$ is the wind speed and $\eta = \tfrac{17}{20}$ is $E$'s fixed rotor efficiency. $E$'s target is $\mathcal{T}_E \equiv \mathit{out} \in [6,7]$. The wake factor $s \in [0,1)$ is encoded as a soft constraint in $\mathcal{B}_{E,S} = \{s = \tfrac{3}{10}\}$, while $\mathcal{B}_{E,H}$ contains the wind speed and $\eta$. Hence $\bar y = (s)$, while $\bar x = (w, \eta, w_{\mathit{eff}}, \mathit{out})$ consists of known values and derived quantities.

		\begin{enumerate}
			\item \textbf{Failure detection.} $E$ evaluates its model under $\mathcal{B}_E$:
			\[
			w_{\mathit{eff}} = 8 \cdot \tfrac{7}{10} = \tfrac{28}{5}, \qquad
			\mathit{out} = \tfrac{28}{5} \cdot \tfrac{17}{20} = \tfrac{119}{25} = 4.76.
			\]
			Since $4.76 < 6$, $\mathcal{M}_E, \mathcal{B}_E \not\models \mathcal{T}_E$ and $E$ initiates explanation generation.

			\item \textbf{Diagnosis.} $E$ encodes $F = \mathcal{M}_E \wedge \mathcal{B}_{E,H} \wedge \mathcal{T}_E \wedge \mathcal{B}_{E,S}$ as a partial MaxSMT problem with hard part $F_H$ and soft part $F_S$:
			\[
			F_H \equiv \mathit{out} = w_{\mathit{eff}} \cdot \eta \land w_{\mathit{eff}} = w \cdot (1 - s) \land w=8 \land \eta = \tfrac{17}{20} \land 0\leq s < 1 \land 6 \leq \mathit{out} \leq 7, \qquad F_S \equiv s = \tfrac{3}{10}
			\]
			The solver returns the restricted MCS $C = \{s = \tfrac{3}{10}\}$.

			\item \textbf{Range characterization.} Removing $C$ from $F$ leaves $s$ free. Eliminating $w_{\mathit{eff}}$ and $\mathit{out}$ (and the known $w$, $\eta$) yields $\varphi$, which simplifies to $\varphi'$:
			\[
			\varphi(s) \;\equiv\; s \geq 0 \;\wedge\; \neg(s \geq 1) \;\wedge\; s \geq -\tfrac{1}{34} \;\wedge\; s \leq \tfrac{2}{17}, \qquad \varphi'(s) \;\equiv\; 0 \leq s \leq \tfrac{2}{17}
			\]
		\end{enumerate}

	\section{Communication in a Chain}
		\label{sec:demonstration}
		In this section, we present an executable proof-of-concept of the explanation exchange on a chain topology, where each component is influenced by exactly one other component and influences exactly one component, excluding the two end components. In a chain, Step~5 recurses: a receiver that cannot satisfy its own target under the received constraint starts Step~1 itself.\footnote{The code for this demonstration is available at: \url{https://github.com/chaotic-fox/lets-talk-this-out}}

		\begin{figure}[t]
			\begin{minipage}[t]{0.33\textwidth}
				\begin{framed}
					\textbf{Turbine (A)}

					\small
Target: $|out - p_t| \leq 1$

Model: $w = X \land \eta \in [0,0.85] \land out=w\cdot\eta$

Influence: $B \rightarrow p_t = 6.5$
				\end{framed}
			\end{minipage}%
			\begin{minipage}[t]{0.33\textwidth}
				\begin{framed}
					\textbf{Controller (B)}

					\small
Target: $|c_p - p_t| \leq 1$

Model: $\emptyset$

Influence: $C \rightarrow c_p = 6.5$
				\end{framed}
			\end{minipage}%
			\begin{minipage}[t]{0.33\textwidth}
				\begin{framed}
					\textbf{Household (C)}

					\small
Target: $\max(x_1+x_2+x_3)$

Model: $a_1 = 2 \land a_2 = 3 \land a_3 = 1.5 \land c_p = a_1\cdot x_1 + a_2\cdot x_2 + a_3\cdot x_3 \land x_1+x_2+x_3\geq 2$
				\end{framed}
			\end{minipage}
			\caption{Simplified energy grid example. $w = X$ denotes the current wind speed. ``Influence: $B \rightarrow p_t = 6.5$'' means that $p_t$ is controlled by $B$ and currently has the value $6.5$. The booleans $x_i$ count as $0/1$ in sums.}
			\label{fig:chain}
		\end{figure}

		Figure~\ref{fig:chain} shows three components in a chain topology. Component A is influenced only by component B, which is influenced only by component C. The turbine's (A) target is to keep the energy within 1 unit of the production goal $p_t$, which is inherited from the controller (B). B's target is to balance the energy production goal $p_t$ and the projected energy consumption $c_p$, which it inherits from a model household (C). C tries to maximize the number of active appliances, under the constraint that it needs to activate at least 2. The appliances 1, 2, and 3 consume 2, 3, and 1.5 energy units each. Non-negativity clauses are omitted; $x_1,x_2,x_3$ are booleans, all other variables are reals.

		At wind speed 7, this system functions without problem, since A can meet the production target for all 3 appliances. However, when the wind speed drops, then first the production goal and finally the appliance schedule need to be adjusted.

		\begin{figure}[t]
			\small
			\begin{minipage}[t]{.5\textwidth}
				\centering
				$\mathbf{w=6}$
\begin{verbatim}
[A] target unsatisfied. MCS = (p_t)
[A] QE -> (0 <= p_t <= 6.1) -> B
[B] target satisfied for p_t = 5.5
\end{verbatim}
				$\mathbf{w=1}$
\begin{verbatim}
[A] target unsatisfied. MCS = (p_t)
[A] QE -> (0 <= p_t <= 1.85) -> B
[B] target unsatisfied. MCS = (c_p)
[B] QE -> (0 <= c_p <= 2.85) -> C
[C] target unsatisfied
\end{verbatim}
			\end{minipage}%
			\begin{minipage}[t]{.5\textwidth}
				\centering
				$\mathbf{w=3}$
\begin{verbatim}
[A] target unsatisfied. MCS = (p_t)
[A] QE -> (0 <= p_t <= 3.55) -> B
[B] target unsatisfied. MCS = (c_p)
[B] QE -> (0 <= c_p <= 4.55) -> C
[C] target satisfied for x1 = 0, c_p = 4.5
[B] target satisfied for p_t = 3.55
\end{verbatim}
			\end{minipage}
			\caption{Abridged logs of the chain explanation exchange}
			\label{fig:logs}
		\end{figure}

		Figure~\ref{fig:logs} shows abridged logs. At wind speed 6, A sends a message to B, requesting a reduction of the production goal. B can accommodate the request without involving C and sends back $p_t = 5.5$. However, at wind speed 3, B cannot accommodate the request on its own, so it sends a request to C, asking it to reduce its projected consumption. C can accommodate the request by turning off one of its appliances, and sends a new value back, $c_p = 4.5$, which B uses to lower the production goal to $p_t = 3.55$. At wind speed 1, every pair of appliances exceeds the requested consumption, so the exchange results in a failure.

		Since each component only has at most one upstream and one downstream neighbor, our explanation generation is sufficient to serve as a minimalistic exchange protocol. As presented by Luckcuck et al.~\cite{luckcuck2026}, in that case each assume $A_i$ and guarantee $G_i$ are connected by the additional rule $G_i \implies A_{i+1}$, which we integrate by adding the explanation $\varphi$ to the receiver's model check. Since a chain has neither cycles nor branches, messages cannot propagate endlessly, therefore any exchange ends after at most $n-1$ messages, where $n$ is the number of components. If a component possesses multiple shared variables, there may also be multiple candidates for $\bar y_C$ in the diagnosis step, and as a result multiple attempts to solve the same conflict if one exchange fails, but since there are finitely many shared variables, there can only be finitely many subsets $\bar y_C$. Termination is guaranteed with either a positive result if a satisfying assignment for every component is found, or negative if no attempt results in one.

	\section{Discussion}
		\label{sec:protocol}
		Section \ref{sec:demonstration} demonstrated that our explanation generation approach is viable for one-dimensional chains of components, but additional mechanisms are needed to guarantee termination in more complex networks. In this section, we discuss concrete challenges and possible future directions for the two main features networks can have: branches and cycles.

		\paragraph{Branching Topologies (tree, star)}
		In reality, components generally have more than a single connection. This has two effects: First, each additional parent increases the number of possible correction sets, which have to be searched for a working solution by a depth-first search. Second, each child may be affected by a change and start its own exchange. To compensate, our protocol must contain mechanisms to search through different candidates to find a solution, as well as consider side effects before committing.
		
		\paragraph{Cyclic Topologies (ring)}
		In cyclic topologies, a clear hierarchy cannot be established, which means we cannot rely on a root node which makes the last decision to guarantee termination. Without precautions, a message may infinitely cycle through the network.
		There are two approaches we are considering at this moment: Firstly, if the protocol could guarantee iterative tightening of constraints, then the exchange will eventually end with either a successful feedback, or a failure when a component determines that tightened constraints exclude any solution (cf. Abstraction Refinement \cite{bobaru2008}). Secondly, if the explanation carries additional information about each involved component, then perhaps we can avoid having to enter a cycle. Once a component would attempt to contact a previously involved component, it could instead directly determine whether the solution is viable from the included constraints.
		
		\paragraph{Conclusion}
		In this research preview paper, we presented a generation method for counterfactual explanations from formal specifications using partial MaxSMT and quantifier elimination. We demonstrated how an exchange of these explanations can allow a simple system of components to organize without the need for a central decision maker. In the future, we would like to embed the explanation generation in a protocol which works in a wider range of topologies as illustrated above. Luckcuck et al.~\cite{luckcuck2026} provide rules and inferences for the branching case, which we would like to adapt. For the cyclic case, they discuss the same difficulties and argue that only partial conclusions can be drawn at this point. In this version, we considered precise knowledge, especially about the environment, but in reality that may not be available. Having only bounds on certain parameters would change the formula used for the range characterization, where $\bar u$ are variables on which we only have bounds:
		\[
		\varphi(\bar y_C) \;\equiv\; \forall \bar u.\, \Big( \mathcal{B}_{E,U}(\bar u) \rightarrow \exists \bar x.\, \big( \mathcal{M}_E \wedge \mathcal{B}_{E,X} \wedge \mathcal{T}_E \wedge (\mathcal{B}_{E,S} \setminus C) \big) \Big)
		\]
			

		\paragraph{Acknowledgement}
			This work has been funded by the Deutsche Forschungsgemeinschaft (DFG, German Research Foundation) project no. 513623283 as part of the Research Training Group CAUSE.

		\paragraph{AI Usage}
			The code for the demonstration was written with use of Anthropic's Claude coding assistant.

		\bibliographystyle{eptcs}
		\bibliography{references_counterfactual_explanations}

\end{document}